\documentstyle[multicol,prb,aps,amssymb,graphicx]{revtex}

\newcount\picturewidth
\begin{document}
\draft

\title{Low field vortex dynamics over seven time decades in a \boldmath
      $\mathbf{Bi_2Sr_2CaCu_2O_{8+\delta}}$ single crystal for temperatures 
      $13 \, {\mathbf{K}} \le T \le 83 \, {\mathbf{K}}$}

\author{M.~Nider\"ost, A.~Suter\cite{suter}, P.~Visani, and A.C.~Mota}
\address{Laboratorium f\"ur Festk\"orperphysik, Eidgen\"{o}ssische 
         Technische Hochschule Z\"urich, \\ 
         8093 Z\"urich, Switzerland}
\author{G.~Blatter}
\address{Theoretische Physik, Eidgen\"{o}ssische 
         Technische Hochschule Z\"urich, \\ 
         8093 Z\"urich, Switzerland}

\date{\today}
\maketitle

\begin{abstract} Using a custom made dc-SQUID magnetometer, we have 
measured the time relaxation of the remanent magnetization 
$M_{\mathrm{rem}}$ of a $\mathrm{Bi_2Sr_2CaCu_2O_{8+\delta}}$ 
single crystal from the fully critical state for temperatures $13 \, 
{\mathrm{K}} \le T \le 83 \, {\mathrm{K}}$.  The measurements cover a 
time window of seven decades $10^{-2} \, {\mathrm{s}} \lesssim t 
\lesssim 10^5 \, {\mathrm{s}}$, so that the current density $j$ can be 
studied from values very close to $j_c$ down to values considerably 
smaller than $j_c$.  From the data we have obtained: (i)~the flux 
creep activation barriers $U$ as a function of current density $j$, 
(ii)~the current-voltage characteristics $E(j)$ in a typical range of 
$10^{-7} \, {\mathrm{V/cm}}$ to $10^{-15} \, {\mathrm{V/cm}}$, and 
(iii)~the critical current density $j_c(0)$ at $T = 0$.  Three 
different regimes of vortex dynamics are observed: For temperatures $T 
\lesssim 20 \, {\mathrm{K}}$ the activation barrier $U(j)$ is 
logarithmic, no unique functional dependence $U(j)$ could be found for 
the intermediate temperature interval $20 \, {\mathrm{K}} \lesssim T 
\lesssim 40 \, {\mathrm{K}}$, and finally for $T \gtrsim 40 \, 
{\mathrm{K}}$ the activation barrier $U(j)$ follows a power-law 
behavior with an exponent $\mu \simeq 0.6$.  From the analysis of the 
data within the weak collective pinning theory for strongly layered 
superconductors, it is argued that for temperatures $T \lesssim 20 \, 
{\mathrm{K}}$ pancake-vortices are pinned individually, while for 
temperatures $T \gtrsim 40 \, {\mathrm{K}}$ pinning involves large 
collectively pinned vortex bundles.  A description of the vortex 
dynamics in the intermediate temperature interval $20 \, {\mathrm{K}} 
\lesssim T \lesssim 40 \, {\mathrm{K}}$ is given on the basis of a 
qualitative low field phase diagram of the vortex state in 
$\mathrm{Bi_2Sr_2CaCu_2O_{8+\delta}}$.  Within this description a 
second peak in the magnetization loop should occur for temperatures 
between $20 \, {\mathrm{K}}$ and $40 \, {\mathrm{K}}$, as it has been 
observed in several magnetization measurements in the literature.  
\end{abstract}

\begin{multicols}{2}
\narrowtext

\section{Introduction}

High temperature superconductors (HTSC) are characterized by large 
values of the Ginzburg-Landau parameter $\kappa = \lambda / \xi$, so 
that most of the $H-T$ phase diagram is dominated by the presence of 
vortices.  Furthermore, the high anisotropy of 
$\mathrm{Bi_2Sr_2CaCu_2O_{8+\delta}}$ (Bi2212) has strong implications 
for the behavior of the flux lattice in the mixed state.  When a 
magnetic field $H$ is applied perpendicularly to the $ab$-planes, the 
vortices can be described as two dimensional 
``pancake-vortices''\cite{clem91} lying in the superconducting 
$\mathrm{CuO_2}$ layers.  These pancake-vortices interact both 
through the interlayer Josephson coupling as well as through
electromagnetic coupling.  Such a layered vortex structure is very 
sensitive to thermal and quantum fluctuations, especially considering 
the small coherence length $\xi$ in the direction parallel to the 
$\mathrm{CuO_2}$ planes.  As a consequence, pinning is relatively weak 
as compared to classical type II superconductors and strong relaxations 
of the magnetization $M$ are 
observed\cite{safar,svendlindh,donglushi,hu,vanderbeek,emmen,zhukov} 
which deviate from a pure logarithmic time dependence.

Since the discovery of the HTSC\cite{mueller}, the theoretical and 
experimental work concerning vortices and their dynamics have been 
strongly intensified.  Those investigations were mainly focused on a 
regime where the current density $j$ is relatively small as compared 
to the critical current density $j_c$.  Only little is 
known\cite{kuepfer,gao,puig} at present regarding the vortex dynamics 
in a regime where the current density $j$ is close to $j_c$.

In the here presented work, we investigate experimentally the low field 
vortex dynamics in a Bi2212 single crystal for magnetic fields 
$\bbox{H}$ perpendicular to the $\mathrm{CuO_2}$ layers.  For this 
purpose we have designed and constructed a dc-SQUID magnetometer with 
high sensitivity and long time thermal stability.  The measurements of 
the relaxation of the remanent magnetization $M_{\mathrm{rem}}$ are 
taken in the temperature interval $13 \, {\mathrm{K}} \le T \le 83 \, 
{\mathrm{K}}$ and cover a time window of seven decades.  The wide 
current range of the experimental data allows a detailed analysis of 
the vortex dynamics within the theoretical vortex creep 
models.\cite{blatter,fisher}  Applying the method of Maley {\it et 
al.}\cite{maley} to the relaxation data, a characteristic functional 
dependence between the activation barrier $U$ and the current density 
$j$ is obtained for the temperature regimes $T \lesssim 20 \, 
{\mathrm{K}}$ and $T \gtrsim 40 \, {\mathrm{K}}$, whereas for 
temperatures between $20 \, {\mathrm{K}}$ and $40 \, {\mathrm{K}}$ the 
$U(j)$-relation is found to depend strongly on temperature.  On the 
basis of a qualitative low field phase diagram\cite{blatter95} of 
Bi2212, an interpretation of the behavior of the vortex dynamics in 
the temperature interval around the crossover 
temperature\cite{zavaS,TT,metlushko94,nideroest} $T \simeq 25 \, 
\mathrm{K}$ is given.

\section{Experimental} \label{Exp}

The measured single crystal is $0.9 \times 1.3 \times 0.05 \, {\rm 
mm}^3$ in size and the critical temperature $T_c$ is 95~K, as 
determined by ac-susceptibility measurements.  The growth procedure as 
well as the transport properties have been described elsewhere.  
\cite{zavaritzky} The experiments are performed in a custom made 
dc-SQUID magnetometer, where the sample remains stationary in the 
pick-up coil during the measurements.  The externally applied magnetic 
field $H$ is supplied by a superconducting coil, working in a 
non-persistent mode.  In order to prevent eddy currents, the 
experimental cell is entirely built out of epoxy resin 
Stycast~1266.\cite{stycast}

During the measuring procedure, the sample is first zero field cooled 
in the residual field of the cryostat ($H^{\mathrm{res}} \simeq 10 \, 
\mathrm{mOe}$) from well above $T_c$ and then stabilized at a fixed 
temperature $T$.  Next, a magnetic field $H$ applied perpendicularly 
to the $ab$-planes is gradually increased from zero to 
$H^{\mathrm{max}}$ before being removed linearly at a rate of 
$\mathrm{9 \, T/s}$.  This fast rate is achieved by shorting the 
superconducting coil ($L \simeq 7 \, \mathrm{H}$) over an extremely 
non-linear resistor.  Measurements of the current in the coil show 
that no discontinuities occur during the removal of the field.  The 
data are taken as soon as the decreasing magnetic field fulfills the 
condition $H < 1 \, \mathrm{Oe}$.  As a time origin for the measured 
data we choose the time at which $H$ starts being removed.  After 
measuring the relaxation of the remanent magnetization 
$M_{\mathrm{rem}}$ for about seven time decades, the sample is heated 
above $T_c$ in order to record its residual magnetization.  The 
maximum values of the initially applied magnetic field 
$H^{\mathrm{max}}$ are shown in Table \ref{ResField} for all the 
measuring temperatures $T$.  The values of $H^{\mathrm{max}}$ are 
chosen so that they are bigger than twice the field needed to achieve 
full flux penetration into the sample.  Table \ref{ResField} further 
contains the values of the residual field $H^{\mathrm{res}}$ along the 
axis of the superconducting coil after cycling the magnetic field from 
zero to $H^{\mathrm{max}}$ and back to zero again.

Due to the high field removal rate $\dot{H} = 9 \, \mathrm{T/s}$, it 
was necessary to perform some controls concerning the initial field 
profile in the sample as well as self-heating effects.  Several field 
removal rates $\dot{H}$ have been tested.  We found that for the field 
removal rates $10^{-2} \, {\mathrm{T/s}} \le \dot{H} \le 9 \, 
{\mathrm{T/s}}$, the measured remanent magnetizations 
$M_{\mathrm{rem}}$ do not show any remarkable difference.  Moreover, 
no significant change in the dynamics of the relaxation of 
$M_{\mathrm{rem}}$ could be detected by increasing the initial field 
values $H^{\mathrm{max}}$ by a factor of 2--3.  From our estimations 
we concluded that the self-heating of the sample due to induction as 
well as to flux flow can be neglected for all temperatures and fields 
of our measurements.

With the described experimental procedure we measured the relaxation 
of the remanent magnetization $M_{\mathrm{rem}}$ for the Bi2212 
crystal from the fully critical state in a temperature range of $13\, 
{\mathrm{K}} \le T \le 83 \, \mathrm{K}$.  Since there is only a small 
uncertainty of the time origin of the creep process ($< 18 \cdot
10^{-3} \, \mathrm{s}$), the initial behavior of the relaxation data 
as a function of time is very well defined so that we were able to 
test the existing vortex creep models over a wide current density 
region starting from values near $j_c$.

\section{Flux dynamics models} \label{TheoAspec}

The main effect of pinning is to allow a flux density gradient to be 
sustained within a type II superconductor.  This is intrinsically 
related to the flow of a macroscopic diamagnetic screening current 
density $j$ that can be expressed, in the continuous approximation, 
through Maxwell's equation ${\bbox{\nabla}} \wedge {\bbox{B}} = 4 \pi 
/ c~ {\bbox{j}}$.  The configuration with a finite flux density 
gradient is metastable and hence is bound to decay.  The dynamics 
arises from the vortex creep motion as a result of thermal 
activation\cite{kim} and quantum tunneling\cite{mota,blatter93} ($T 
\lesssim 5 \, {\mathrm{K}}$).  For a geometry where 
${\bbox{B}}~\|~\hat{z}$ and $\bbox{j}~\|~\hat{y}$, the Maxwell 
equations together with the condition of flux conservation lead to the 
non-linear diffusion equation\cite{beasly,brandt91}

\begin{equation}\label{DiffEq}
   \frac{\partial j}{\partial t} = \frac{c}{4 \pi} \, 
   \frac{\partial^2}{\partial x^2} (v B). 
\end{equation}

Anderson\cite{anderson} postulated that the velocity of the vortices, as a 
consequence of thermal activation over the pinning barrier $U(j)$, be 
given by
     
\begin{equation}\label{velocity}
  v = v_o(j) \, \exp\left( -\frac{U(j)}{k_B T} \right),
\end{equation}

\noindent where $v_o(j)$ is the mean vortex velocity and can be 
expressed as $v_o(j) = l(j) / \tau_o$, where $l(j)$ is the mean 
hopping length and $\tau_o$ the is inverse attempt frequency.  For a 
situation where $v_o(j)$ is independent of $j$, the diffusion 
equation (\ref{DiffEq}) can be transformed into

\begin{equation}\label{CurrentEq}
  \frac{\partial j}{\partial t} \simeq -\frac{j_c}{\tau_o} \exp\left( -\frac{U(j)}{k_B T} \right).
\end{equation}              

\noindent As discussed by Geshkenbein and Larkin\cite{geshkenbein}, equation 
(\ref{CurrentEq}) can be solved within logarithmic accuracy, yielding
     
\begin{table} 
\caption{Maximum applied magnetic field $H^{\mathrm{max}}$ for different
         measuring temperatures $T$ and values of the residual field $H^{\rm 
         res}$, which is due to the flux remaining trapped in the 
         superconducting coil after the removal of the external field $H$.  The 
         residual field of the cryostat is about $10 \, \mathrm{mOe}$ in 
         opposite direction to the applied magnetic field.}
\protect\label{ResField}         
\begin{tabular}{cccccc}
     $T \, {(\mathrm{K})}$ & 13 & 15 -- 27 & 30 -- 40 & 50 & $\ge 60$ \\ 
     \hline $H^{\rm max} {(\mathrm{Oe})}$ & 1500 & 1000 
     & 500 & 300 & $< 100$ \\ \hline $H^{\rm res} 
     {(\mathrm{mOe})}$ & $710 \pm 50$ & $480 \pm 50$ & $60 \pm 10$ & $20 
     \pm 10$ & $-10 \pm 10$ \\
\end{tabular}  
\end{table}

\begin{equation}\label{Ut}
   U(j(t)) \simeq k_B T \ln\left( 1 + \frac{t}{t_o} \right),
\end{equation}     

\noindent where $t_o = k_B T \tau_o / j_c |\partial_j U|$ is a time 
scaling factor.  Once the functional dependence between the pinning barrier 
$U$ and the current density $j$ is known, the time dependence of $j$ is 
simply determined by the inversion of (\ref{Ut}).
     
On approaching the critical current density $j_c$, the effective pinning 
barrier vanishes and one can write
     
\begin{equation}\label{U_GKA}
  U(j \to j_c) \simeq U_c \left( 1 - \frac{j}{j_c} \right)^\alpha .
\end{equation}

\noindent Comparing equations (\ref{Ut}) and (\ref{U_GKA}), the 
following time dependence of $j$ is obtained:
     
\begin{equation}\label{j_GKA}
  j(t) \simeq j_c \left[ 1 - \left\{ \frac{k_B T}{U_c} \ln\left( 1 + 
  \frac{t}{t_o} \right) \right\}^{1 / \alpha} \right],~~~~j \to j_c,
\end{equation}     

\noindent which maps to the original formulation of 
Anderson\cite{anderson} for $\alpha = 1$.
     
In the above derivation it is assumed that the current densities $j$ 
be close to $j_c$.  This is a good assumption for conventional type II 
superconductors.  Further theoretical considerations are necessary to 
describe the strongly decaying current densities in HTSC, for which 
values of $j$ much smaller than $j_c$ are reached already at 
laboratory times.  For the HTSC in the limit of small currents, the 
weak collective pinning theory\cite{blatter} (WCPT) as well as the 
vortex glass theory\cite{fisher}, predict an activation barrier that 
diverges algebraically for vanishing currents:

\begin{equation}\label{U_jsmall}
  U(j) \simeq U_c \left( \frac{j_c}{j} \right)^\mu.
\end{equation}    
      
\noindent Inserting the relation (\ref{U_jsmall}) into equation 
(\ref{Ut}) the following non purely logarithmic time dependence of the 
current density $j$ is obtained:
     
\begin{equation}\label{jt_jsmall}
   j(t) \simeq j_c \left[ \frac{k_B T}{U_c} \ln\left( \frac{t}{t_o} \right) \right]^{-1/\mu},~~~~j \ll j_c.
\end{equation}           

In order to find a more general formula, (\ref{jt_jsmall}) and 
(\ref{j_GKA}) (we assume $\alpha = 1$) can be interpolated with the 
following expression

\begin{equation}\label{j_interpol}
  j(t) \simeq j_c \left[ 1 + \mu \frac{k_B T}{U_c} \ln\left( 1 + \frac{t}{t_o} 
       \right) \right]^{-1/\mu},
\end{equation}

\noindent and the corresponding activation barrier is (see equation 
 (\ref{Ut}))
     
\begin{equation}\label{U_interpol}
   U(j) \simeq \frac{U_c}{\mu} \left[ \left( \frac{j_c}{j} \right)^\mu - 1 \right].
\end{equation}

Within the \textit{single vortex} pinning regime the exponent 
$1/\mu~\approx~7$ is large, such that for $\mu \, k_B T / U_c \, 
\ln\left( 1 + t/t_o \right) \ll 1$ expression (\ref{j_interpol}) can be 
approximated by

\begin{equation}\label{PowerLaw}
   j(t) \simeq j_c \left( 1 + \frac{t}{t_o} \right)^{- k_B T / U_c},
\end{equation}
     
\noindent with a logarithmic potential 

\begin{equation}\label{U_log}
   U(j) \simeq U_c \ln(j_c/j).
\end{equation}

Notice that within the WCPT, the divergence in the potentials 
(\ref{U_jsmall}), (\ref{U_interpol}), and (\ref{U_log}) at low current 
densities $j$ is related to the observation that the activated motion 
of vortices involves hops of larger vortex segments/bundles over 
longer distances.  The elastic energy cost will therefore grow with 
decreasing $j$.  This is no longer the case for the point-like 
pancake-vortices for which no extra deformation energy is needed in 
order to overcome the pinning barrier for decreasing current 
densities.  For strongly layered superconductors within the 
\textit{single pancake} creep regime the activation barrier $U(j)$ is 
therefore expected to saturate.  However, using the concept of 
variable-range hopping\cite{mott,shklovskii} (VRH) it has been 
argued\cite{blatter} that, for decreasing current densities $j$, 
pancake-vortices still couple into a 2D elastic manifold.  As a matter 
of fact, due to the randomness in the energies of the metastable 
state, pancakes will hop over larger distances as the current density 
$j$ decreases.  Such a large hopping distance $u$ leads to a large 
shear interaction energy $c_{66} d u^{2}$ between pancakes ($c_{66}$ 
is the shear modulus and $d$ the interlayer distance).  As a 
consequence, for a large enough hopping distance $u$, the 
pancake-vortex will start to couple to its neighbours.  Thus, the 
vortex system is expected to first go through a VRH regime, which is 
followed by a 2D collective creep regime\cite{vanderbeek} at still 
lower current densities.

In the above treatment of the flux dynamics models we have 
considered current densities $j$ flowing inside a superconductor, 
whereas from the experiment we obtain spatially averaged values of the 
magnetization $M$.  In the case of an infinite slab parallel to the 
applied magnetic field $H$, the dependence between $M$ and $j$ has 
been described by Bean.\cite{bean}  Recently Gurevich and 
Brandt\cite{brandt94} obtained an asymptotic solution for the 
non-linear diffusion equation (\ref{DiffEq}) describing flux creep in 
strips and disks starting from a barrier as given in formula 
(\ref{U_interpol}).  It turns out that, despite the particular field 
distribution for these sample geometries, the current density $j$ can 
still be considered as constant throughout the sample at a given time 
$t$.  It follows that the magnetization $M$, which is given 
by

\begin{equation} 
   \bbox{M}(t) = \frac{1}{V} \cdot \frac{1}{2 c} \,
   \int \bbox{r} \wedge \bbox{j}(\bbox{r},t) \,dV,
\end{equation}

\noindent for a disk-like geometry and for a constant current density 
$\bbox{j}(\bbox{r},t) = j(t) \, \bbox{e_\phi}$, can be expressed as

\begin{equation} 
   {|\bbox{M}(t)|} = j(t) \cdot \frac{1}{V} \cdot \frac{1}{2 c} \,
    \int {|\bbox{r} \wedge \bbox{e_\phi}|} \,dV,
\end{equation}

\noindent where the integration over the geometrical factor leads to

\begin{equation}\label{MpropJ}
   M(t) \simeq \frac{R}{3c} \cdot j(t),
\end{equation}

\noindent with $R$ being the sample radius.  For the case of disks 
(strips) the well known Bean model relationship for an infinite 
cylinder (infinite slab) in the fully critical state is therefore 
still a valid approximation.

\section{Vortices in strongly layered superconductors} 
  \label{Vortex lattice}

For the considerations given in this Section concerning the vortex 
lattice in coupled superconducting layers we will closely follow the 
approach of Refs.\ \onlinecite{blatter} and \onlinecite{koshelev93}.  
Within weak collective pinning theory the size of the correlated 
regions (Larkin domains) is determined by the balance between 
deformation energy and pinning energy.  In terms of length scales, the 
volume forming the Larkin domain is given by the pinning correlation 
lengths $R_c$ and $L_c$ in the direction perpendicular and parallel to 
the magnetic field, respectively.  Through the study of the relative 
magnitude of the deformation and the pinning energy of a vortex 
lattice in coupled superconducting layers, it is possible to determine 
the size of the correlated regions as a function of temperature and 
field.

For a magnetic field $H$ perpendicular to the superconducting layers, 
a vortex lattice has three relevant energy scales, namely the tilt 
energy $U_{\rm tilt} \approx c_{44}(R_c) \cdot(r_p/ L_c)^2 
R_c^2 L_c$, the shear energy $U_{\rm shear} \approx c_{66} \cdot 
(r_p/ R_c)^2 R_c^2 L_c$, and the pinning energy $U_{\rm pin} \approx 
(\gamma \xi^4 R_c^2 L_c / r_p^2 a_o^2)^{1/2}$, with $c_{44}$ being the 
dispersive tilt modulus, $c_{66}$ the shear modulus, $r_p(T)$ the 
range of the pinning force, $a_o$ the intervortex spacing, and 
$\gamma$ the disorder strength (where a short-scale correlated 
disorder potential has been assumed $\langle U_{\rm pin}(r), U_{\rm 
pin}(r') \rangle = \gamma \delta(r-r')$).  Depending on the relative 
magnitude of these energies, one can distinguish four possible pinning 
regimes: (1)~independently pinned vortex pancakes (0D pinning regime: 
$U_{\rm pin} > U_{\rm tilt}, U_{\rm shear}$), (2)~independently pinned 
vortex lines (1D pinning regime: $U_{\rm tilt} > U_{\rm pin} > U_{\rm 
shear}$), (3)~a 2D collectively pinned state in which the 2D vortex 
lattices in the layers are pinned independently from each other 
($U_{\rm shear} > U_{\rm pin} > U_{\rm tilt}$), and (4)~a 3D 
collectively pinned state ($U_{\rm tilt}, U_{\rm shear} > U_{\rm 
pin}$).

According to Refs.\ \onlinecite{blatter} and \onlinecite{blatter95}, 
for temperatures $T < T_{0} \approx (U_{\rm pc}^2 \, E_{\rm 
pc})^{1/3}$ and fields $B < B_{02} \approx 10 \, \Phi_o/(2\pi\xi^2) \, 
(j_c(0)/j_o)$, where $T_{ 0}$ is a few tens of Kelvins and $B_{02}$ is 
a few Teslas, the dominant energy scale for the strongly anisotropic 
Bi2212 is the pinning energy $U_{\rm pin}$ (where $U_{\rm pc} \simeq 
\varepsilon_o d (j_c / j_o)$, $E_{\rm pc} \approx \varepsilon_o d \, ( 
\xi / \lambda )^2$, $\varepsilon_o = (\Phi_o / 4 \pi \lambda)^2$, 
$\Phi_o = h c/2 e$, $d$ is the interlayer distance, and $j_o$ the 
depairing current density).  The $B$-$T$ phase diagram for this region 
is therefore characterized by 0D pinning.  On the other hand, for 
temperatures $T > T_{0}$ the collective pinning length $L_c$ and the 
collective pinning radius $R_c$ both grow very fast due to thermal 
depinning.  This implies that for temperatures $T \gtrsim 20 \, 
{\mathrm{K}}$ the size of the Larkin domains becomes large giving rise 
to a 3D pinning regime\cite{blatter95}.  At high fields $B > B_{23}$, 
a crossover to a 2D collective pinning region is 
predicted\cite{blatter} when the shear energy outweighs the tilt 
energy.

\begin{figure}
\includegraphics[width=0.95\linewidth]{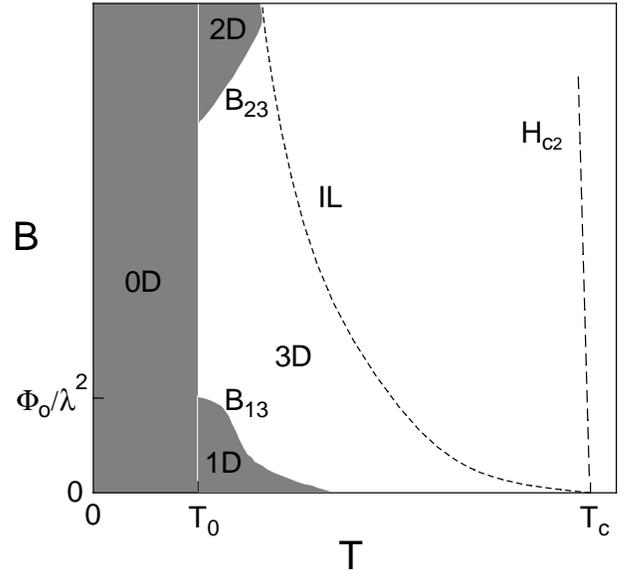}
 \caption{Qualitative low field phase diagram of the vortex state in 
          Bi2212 for magnetic fields $H$ perpendicular to the superconducting 
          layers.  The differently shaded areas in the Figure represent the 
          following pinning regimes: 0D, individually pinned pancake-vortices; 
          1D, individually pinned vortex lines; 2D, collectively pinned state 
          in which 2D lattices of pancake-vortices in the layers are pinned 
          independently from each other; 3D, collectively pinned vortex bundles.  
          $B_{13}$ ($B_{23}$) represent the fields at which the 1D (2D) regime 
          crosses over to the 3D regime. $T_{0}$ is a crossover temperature  
          terminating single pancake pinning.  A sketch of the irreversibility  
          line IL and of the upper critical field $H_{c2}$ is also given.}
\label{PicPD_I}
\end{figure}

Finally, since the relaxation measurements of the remanent 
magnetization $M_{\rm rem}$ presented in this work are performed in 
the ``field off'' state, we need to discuss the very low field regime.  
At fields $B > \Phi_o / \lambda^2$, the shear modulus $c_{66}$ 
has a linear dependence in $B$, whereas at low fields ($B < \Phi_o / 
\lambda^2$), $c_{66}$ decreases exponentially\cite{blatter95,labusch}

\begin{equation} \label{c66}
  c_{66} \approx \left\{ \begin{array}{l@{\qquad}l}
     \displaystyle \frac{\varepsilon_o}{\lambda^2} \,
                   \left(\frac{B \lambda^2}{\Phi_o}\right)^{1/4} \, 
                   e^{-\sqrt{\Phi_o / B \lambda^2}} \ ,& 
                   B < \Phi_o / \lambda^2 \ , \\
     \\    
     \displaystyle \frac{\varepsilon_o B}{4 \Phi_o}\ ,& 
                   B > \Phi_o / \lambda^2 \ , \\
                         \end{array} \right. 
\end{equation}

\noindent where $\lambda$ is the penetration depth.  As a consequence, 
also the shear energy $U_{\rm shear}$ decreases exponentially for 
fields $B < \Phi_o / \lambda^2$.  This means that for temperatures $T 
> T_{0}$ and small enough magnetic fields ($B < B_{13}$) a 1D pinning 
regime occurs.  Fig.\,\ref{PicPD_I} shows a qualitative map of the low 
field pinning regimes of Bi2212 resulting from these considerations.

\begin{figure}
\includegraphics[width=0.95\linewidth]{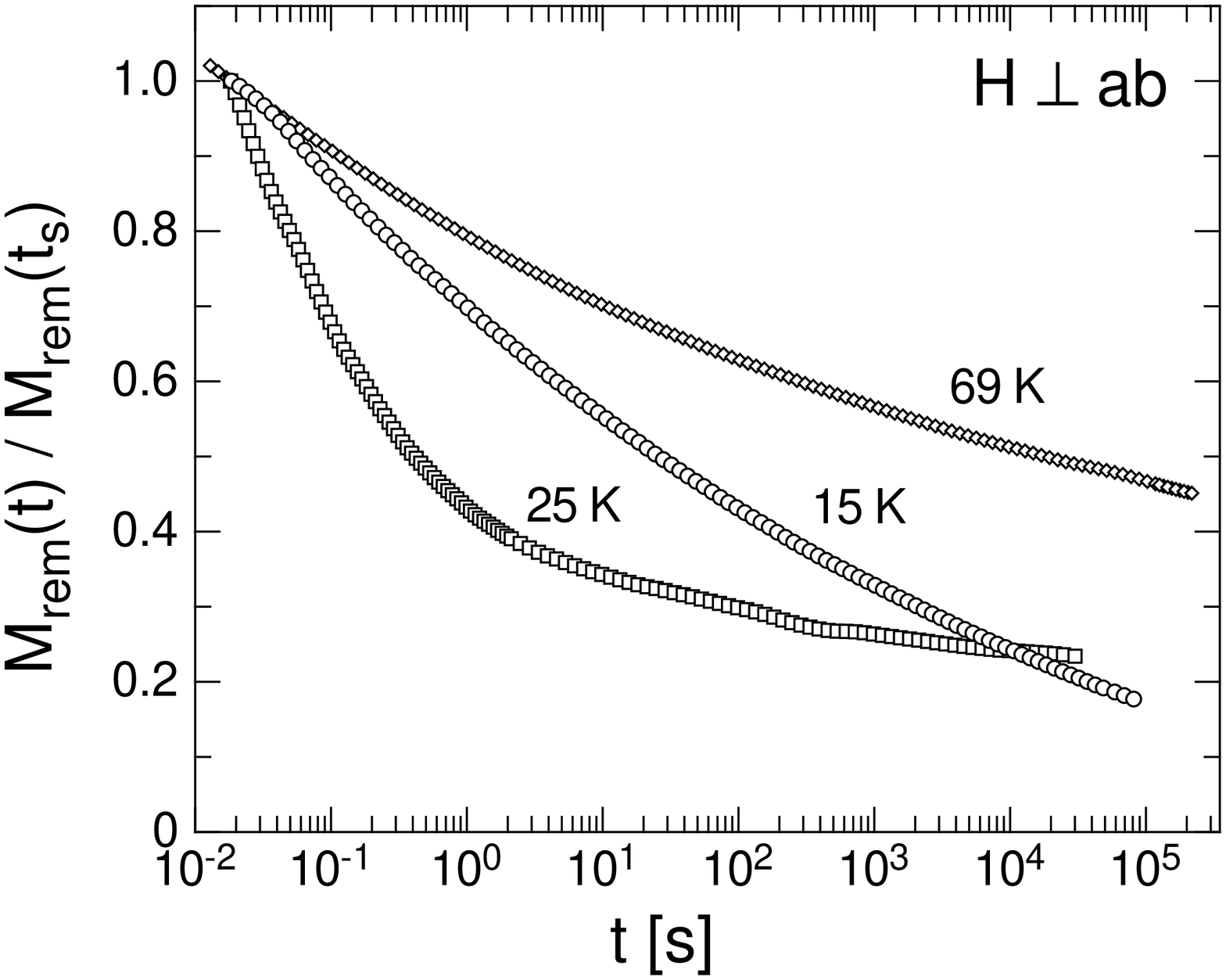}
\caption{Normalized remanent magnetization vs.  time, measured after 
         cycling the sample in an external magnetic field $H$.  In parenthesis 
         the values of the maximal cycling fields $H^{\rm max}$ are given: 
         \protect\newline {\Large $\circ$}~($H^{\rm max} = 1 \, 
         {\mathrm{kOe}}$) at 15 K, $\Box$~($H^{\rm max} = 1 \, {\mathrm{kOe}}$) 
         at 25 K, and {\Large $\diamond$}~($H^{\rm max} = 40 \, {\mathrm{Oe}}$) 
         at 69 K.\protect\newline The time origin is given by the instant, when 
         the externally applied magnetic field $H$ starts being decreased and 
         $t_s \simeq 18 \cdot 10^{-3} \, \mathrm{s}$ is the time, when the 
         first point of the relaxation of $M_{\rm rem}$ is taken.}
\label{PicMremNorm}
\end{figure}

\section{Experimental Results and Analysis}\label{ExpAn}

In Fig.\,\ref{PicMremNorm} we illustrate the time dependence of the 
remanent magnetization $M_{\rm rem}$ with a typical set of data.  A 
non-logarithmic behavior\cite{nideroest} is observed at all 
temperatures.  Notice that at $25 \, \mathrm{K}$, where a sharp drop 
in the relaxation rate $S = -\partial\ln M_{\rm rem}/\partial\ln t$ 
has been previously reported\cite{zavaS,TT,metlushko94}, the remanent 
magnetization $M_{\rm rem}$ decays extremely fast in the first few 
seconds after the removal of the external field $H$.  This is also 
seen in Fig.\,\ref{PicjT}, where the current density $j$ (as obtained 
from formula (\ref{MpropJ})) is plotted as a function of temperature 
for the times $t_s \simeq 18 \cdot 10^{-3} \, \mathrm{s}$, $t_1 = 1 \, 
{\mathrm{s}}$ and $t_2 = 10^{4} \, {\mathrm{s}}$.  The data taken at 
the starting time $t_s \simeq 18 \cdot 10^{-3} \, \mathrm{s}$ (empty 
circles) suggest the presence of only two regimes of vortex dynamics, 
separated by a crossover at $T \simeq 30 \,\mathrm{K}$.  Both regimes 
are accurately described by an exponential temperature dependence but 
with different slopes $d \ln{j}/dT$.  However, at longer times $t 
\gtrsim 1 \, \mathrm{s}$ (filled circles and empty diamonds) the 
existence of a third regime for temperatures between $20 \, 
\mathrm{K}$ and $40 \, \mathrm{K}$ becomes evident.  This third regime 
is characterized by very particular vortex dynamics and will be 
referred to as the ``intermediate regime''.  We will discuss these 
temperature regimes separately and distinguish them as follows: a low 
temperature regime for $T \lesssim 20 \, \mathrm{K}$, an intermediate 
regime for $20 \, {\mathrm{K}} \lesssim T \lesssim 40 \, {\mathrm{K}}$ 
and a high temperature regime for $T \gtrsim 40 \, \mathrm{K}$.  For 
each regime we determine the activation barrier $U(j)$ by means of the 
method of Maley {\it et al.}\cite{maley} Once the functional 
dependence of $U(j)$ is obtained, an analysis of the time evolution of 
the current density $j$ is given.

\subsection{Low Temperature Regime (\boldmath$T \protect\lesssim 20 \, \mathrm{K}$)}

As shown by Maley {\it et al.} it is possible to determine 
the activation barrier for vortex motion $U(j)$ directly from the 
relaxation data $j(t)$.  Starting from equation (\ref{CurrentEq}), one 
obtains

\begin{equation} \label{Uj}
   U(j) \simeq - k_B T \, \ln\left|s \, \frac{\partial j}{\partial t} \right| 
               + k_B T \, \ln\left|s \, \frac{j_c}{\tau_o} \right|,
\end{equation}         

\noindent where the term $k_B T \, \ln\left|s \, j_c /\tau_o \right|$ 
is independent of $j$, and $s = 1 \, \mathrm{cm^2 s/A}$.  Plotting the 
expression $- k_B T \, \ln\left|s \, {\partial j}/{\partial t} 
\right|$ as a function of current density at different temperatures 
$T$, a set of curves is found which are vertically shifted with 
respect to each other.  For a temperature interval where the 
functional dependence between the activation barrier $U$ and the 
current density $j$ is essentially temperature independent, this shift 
is given by the term $a \Delta T$, where $a \simeq \ln\left| s \, 
j_c/\tau_o \right|$ is a constant, and $\Delta T = T_2 - T_1$ is the 
temperature difference between two considered curves.  Combining the 
data measured at different temperatures $T$, the activation barrier $U(j)$ 
is obtained over a wide current density range.

\begin{figure}
\includegraphics[width=0.95\linewidth]{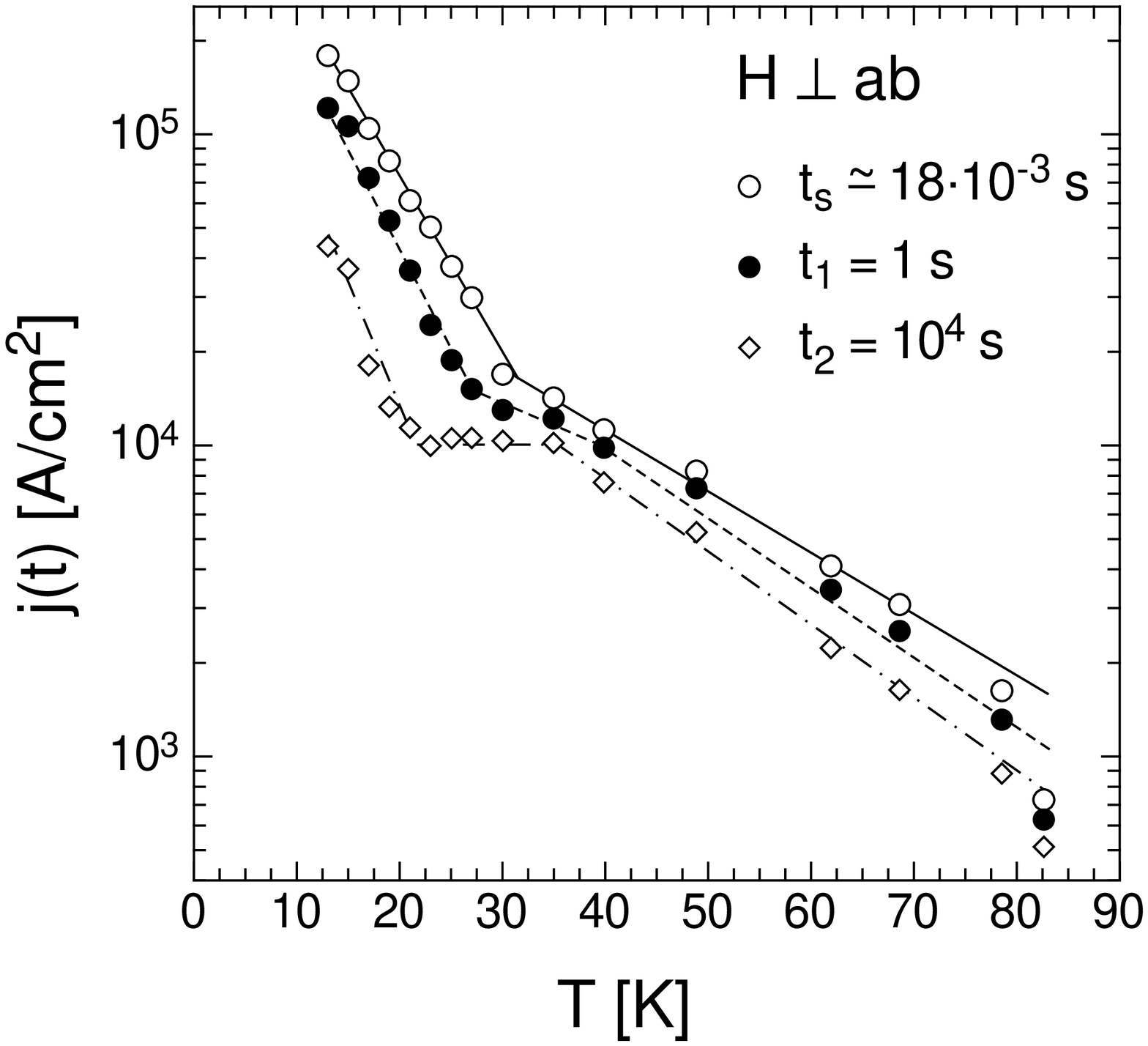}
\caption{Current density $j$ as a function of temperature for 
         different times $t$: \protect\newline {\Large $\circ$}
         starting time $t_s \simeq 18 \cdot 10^{-3}\, \mathrm{s}$, {\Large 
         $\bullet$} $t_1 = 1 \, \mathrm{s}$, and {\Large $\diamond$} 
         $t_2 = 10^4 \, \mathrm{s}$.  \protect\newline The lines serve as 
         guides to the eyes.}
\label{PicjT}
\end{figure}

\begin{figure}
\includegraphics[width=0.8\linewidth]{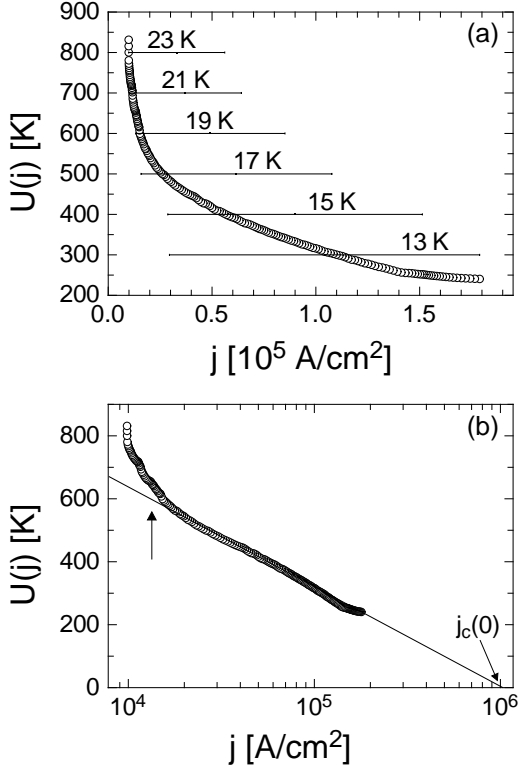}
 \caption{(a) Flux creep activation barrier for temperatures $13 \, 
         {\mathrm{K}} \leq T \leq 23 \, {\mathrm{K}}$ as determined from the 
          magnetic relaxation data by the method of Maley {\it et al.\/} (the 
          constant used for matching the curves is $a = 26 \pm 1$).  The 
          horizontal segments represent the current windows as obtained from the 
          data at a fixed temperature $T$.  \protect\newline (b) The same data 
          in a semi-logarithmic graph.  The line is a fit for temperatures up to 
          $T = 19 \, \mathrm{K}$ (indicated by the arrow) with a logarithmic 
          potential of the type $U(j) \simeq U_{c}\ln(j_{c}/j)$.  From the fit one 
          finds the $T = 0$ critical current density $j_c(0) \simeq (1.0 \pm 
          0.3) \cdot 10^6 \, {\mathrm{A/cm^2}}$.}
\label{PicULow}
\end{figure}

For temperatures $T \lesssim 20 \, {\mathrm{K}}$, the data obtained 
from the expression $- k_B T \, \ln\left|s \, {\partial j}/{\partial 
t} \right|$ at different temperatures $T$ can be accurately mapped 
onto a common curve using a single constant $a$.  The obtained 
potential $U(j)$ is shown in Fig.\,\ref{PicULow}.  It is interesting 
to observe in Fig.\,\ref{PicULow}(a), that the data measured at a fixed 
temperature $T$ (marked by horizontal segments) do overlap over wide 
regions of current.  As seen in Fig.\,\ref{PicULow}(b), the potential 
$U(j)$ is proportional to the logarithm of the current density $j$ 
over a wide current region.  This is in good agreement with previous 
relaxation measurements by van der Beek {\it et al.}\cite{vanderbeek} 
and by Emmen {\it et al.}\cite{emmen}, who found a logarithmic 
dependence of $U(j)$ for temperatures $4 \, {\mathrm{K}} \lesssim T 
\lesssim 17 \, {\mathrm{K}}$.  The deviation from the logarithmic 
behavior at temperatures $T > 19 \, \mathrm{K}$ is attributed to the 
influence of the approaching intermediate regime.

For temperatures $T \lesssim 19 \,\mathrm{K}$, a fit of the measured 
potential $U(j)$ with the logarithmic activation barrier (\ref{U_log}) 
leads to the following parameters: $U_c \simeq 140 \, \mathrm{K}$ and 
the extrapolated critical current density $j_c(T=0) \simeq 1 \cdot
10^6 \, {\mathrm{A/cm^2}}$.  The value of $j_c(T=0)$ is very close to 
the values found in the literature\cite{vanderbeek,emmen} (taking into 
account the considered proportionality factors between $M$ and $j$).

As discussed in Section \ref{TheoAspec}, the activation barrier $U(j)$ 
is expected to be logarithmic within the single vortex pinning regime.  
Since the measured potential $U(j)$ is indeed logarithmic, this would 
suggest that for temperatures $T \lesssim 20 \, \mathrm{K}$ vortex 
strings are pinned individually.  However, a simple estimate of the 
collective pinning length along the $c$-axis $L_c^c \simeq \varepsilon 
\xi (j_o / j_c)^{1/2}$, where $\varepsilon$ is the anisotropy factor 
and $j_o$ is the depairing current density, shows that, for the 
parameters of Table \ref{FitPara} and \ref{TheoPara}, $L_c^c \simeq 
2\,{\mathrm{\AA}} < d = 15\,{\mathrm{\AA}}$.  This means that, for 
temperatures $T \lesssim 20 \, \mathrm{K}$ and low enough magnetic 
fields, pancake-vortices placed on different superconducting layers 
are pinned independently indicating the presence of a single pancake 
pinning regime.  A more detailed discussion of the low temperature 
activation barrier will be given in Section \ref{discussion}.

\begin{figure}
\includegraphics[width=0.95\linewidth]{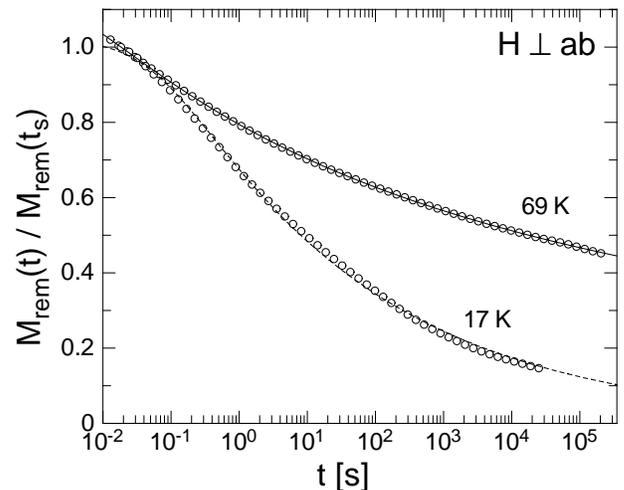}
 \caption{Normalized remanent magnetization vs. time for temperatures 
          $T = 17 \, \mathrm{K}$ and $T = 69 \, \mathrm{K}$.  The lines are fits 
          according to formula (\protect\ref{PowerLaw}) for the 17~K 
          data and formula (\protect\ref{j_interpol}) for the 69~K data.}
\label{PicDecayFit}
\end{figure}

We can crosscheck the result for the barrier as obtained via the Maley 
analysis making use of equations (\ref{PowerLaw}) and (\ref{MpropJ}).  
A typical fit to the data measured at temperatures $T \lesssim 19 \, 
\mathrm{K}$ is shown in Fig.\,\ref{PicDecayFit}, confirming the 
logarithmic dependence $U(j) \simeq U_{c}\ln(j_{c}/j)$.  The resulting 
fitting parameters are the following: $U_c \simeq 140 \, \mathrm{K}$, 
$t_o \simeq 3 \cdot 10^{-2} \, \mathrm{s}$ and values of $j_c(T)$ 
about 5\% above the values shown in Fig.\,\ref{PicjT} for $t_s \simeq 
18 \cdot 10^{-3} \, \mathrm{s}$.

Finally, we point out that for temperatures $T \lesssim 20 \, 
{\mathrm{K}}$ the values of the pinning potential $U(j)$ and of the 
extrapolated critical current density $j_c(T=0)$ are both in good 
agreement with the results in the literature, usually obtained in the 
``field on'' mode at much slower field ramping rates $\dot{H}$.

\subsection{Intermediate Regime (\boldmath$20 \, {\mathbf{K}} 
\protect\lesssim T \protect\lesssim 40 \, {\mathbf{K}}$)} 
            
\begin{figure}
\includegraphics[width=0.95\linewidth]{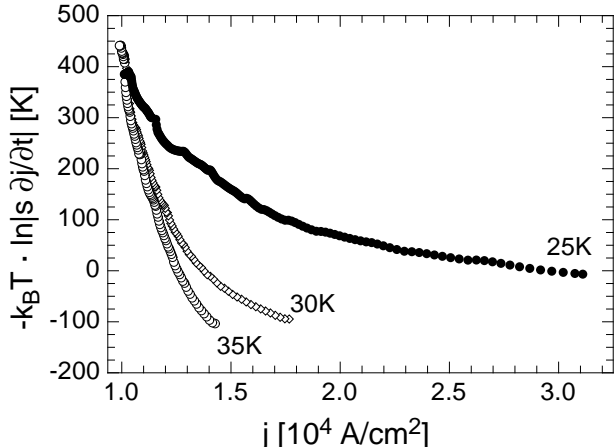}
\caption{Flux creep activation barrier vs. current density $j$ for temperatures
         $23 \, {\mathrm{K}} \leq T \leq 35 \, {\mathrm{K}}$. The vertical axis 
         is only defined up to a constant value. For this temperature 
         regime the curves are strongly tilted with respect to each other
         and cannot be ``glued'' onto a common curve anymore.}
\label{PicU23to35}
\end{figure}

In order to find the activation barrier $U(j)$ for temperatures $20 \, 
{\mathrm{K}} \lesssim T \lesssim 40 \, {\mathrm{K}}$, the relaxation 
data are again evaluated with the method of Maley {\it et 
al.}  The results obtained with help of equation 
(\ref{Uj}) for different temperatures $T$ are shown in 
Fig.\,\ref{PicU23to35}.  We observe that the curves are strongly 
tilted with respect to each other and it is not possible to obtain a 
unique smooth curve by simply shifting the data obtained at different 
temperatures $T$ along the vertical axis.  Thus, within the 
temperature range $20 \, {\mathrm{K}} \lesssim T \lesssim 40 \, 
{\mathrm{K}}$, we cannot find a unique temperature independent 
functional relation between $U$ and $j$ following the above approach.  
A qualitative interpretation of the vortex dynamics in this 
temperature regime will be given in Section \ref{discussion}.

\subsection{High Temperature Regime (\boldmath$40 \, {\mathbf{K}} 
\protect\lesssim T 
            \protect\lesssim 83 \, {\mathbf{K}}$)}

\begin{figure}
\includegraphics[width=0.8\linewidth]{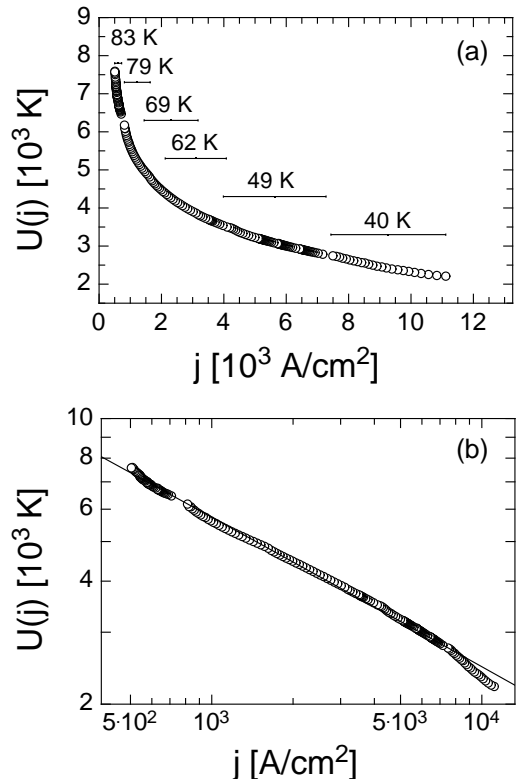}
 \caption{(a) Flux creep activation barrier for $40 \, {\mathrm{K}} 
          \leq T \leq 83 \, {\mathrm{K}}$ as determined by the method of Maley 
         {\it et al.\/} (with $a = 62 \pm 1$) from the magnetic relaxation 
          data.  The horizontal segments represent the current windows as 
          obtained from the data at a fixed temperature $T$.  \protect\newline 
          (b) The same data in a double-logarithmic graph.  The line is a fit 
          for temperatures $T \ge 62 \, \mathrm{K}$, with a power-law potential 
          as given in formula (\protect\ref{U_interpol}).}
 \label{PicMaley_High_Temp}
\end{figure}

For temperatures $T \gtrsim 40 \, {\mathrm{K}}$, the activation 
barrier $U(j)$ is found with the same method that has been applied for 
the low temperature regime using a single constant $a$.  The 
resulting barrier $U(j)$ is shown in Fig.\,\ref{PicMaley_High_Temp}.  
From the double logarithmic plot of Fig.\,\ref{PicMaley_High_Temp}(b), 
we observe that the activation barrier $U(j)$ follows a power-law 
behavior over a wide current range.  Fitting this potential with 
formula (\ref{U_interpol}) for temperatures $62 \, {\mathrm{K}} \le T 
\le 83 \, {\mathrm{K}}$, we find the values $U_c \simeq 1000 \, 
\mathrm{K}$ and $\mu \simeq 0.6$.

According to Ref.\ \onlinecite{blatter}, a power-law potential with 
the form of (\ref{U_interpol}) leads to a time dependence of the 
current density $j$ as given by the interpolation formula 
(\ref{j_interpol}).  As one can see from the solid line of the fit to 
the $T = 69 \, {\mathrm{K}}$ data in Fig.\,\ref{PicDecayFit}, the time 
dependence of the current density $j$ is very well described by the 
interpolation formula (\ref{j_interpol}).  The fitting parameters 
confirm the results previously obtained for the barrier and can be 
summarized as follows (see also Table \ref{FitPara}): $U_c \simeq 1000 
\, \mathrm{K}$, $\mu \simeq 0.6$, $t_o \simeq 3 \cdot 10^{-3}\, 
\mathrm{s}$, and values of $j_c(T)$ about 5\% above those obtained 
from Fig.\,\ref{PicjT} at $t_s \simeq 18 \cdot 10^{-3} \, \mathrm{s}$.  
According to weak collective pinning theory, an exponent $\mu \simeq 
0.6$ indicates a regime of large 3D bundle pinning.

The high temperature data have been analysed considering a constant 
current density $j$ inside the sample

\begin{figure}
\includegraphics[width=0.95\linewidth]{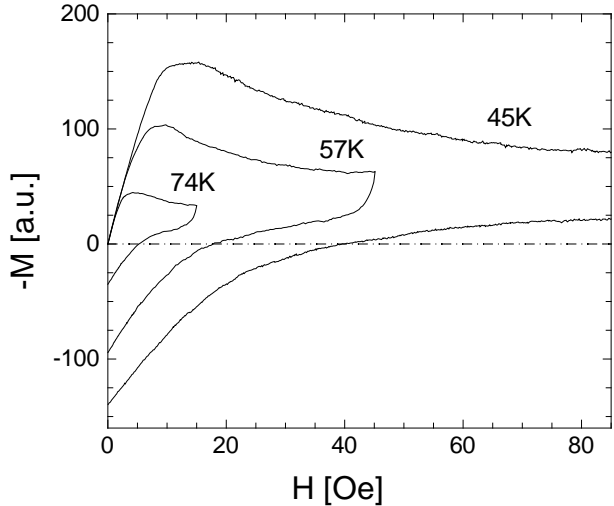}
\caption{Magnetization curves at different temperatures $T$ for the Bi2212 
         crystal. The descending branches of the loops do not have 
         zero magnetization so that we conclude that 
         Bean-Livingston surface barriers are only of secondary importance.}
\label{PicMag}
\end{figure}

\noindent as assumed in the Bean model 
(see formula (\ref{MpropJ})).  We argue that, for the present 
measurements of the relaxation of the remanent magnetization 
$M_{\mathrm{rem}}$, the contributions of pinning due to potential 
barriers arising from surface 
effects\cite{bl,evetts,clem74,burlachkov} and sample geometry 
\cite{zeldov94,zeldov95} have only a secondary effect as compared to 
the contributions of bulk pinning.  As a matter of fact, if surface 
barriers were the only mechanism responsible for the irreversible 
behavior, the magnetization curves would be characterized by 
\textit{zero magnetization} on the descending branch of the 
loop.\cite{evetts,clem74,burlachkov}  Fig.\,\ref{PicMag} shows three 
magnetization cycles of the Bi2212 crystal measured at different 
temperatures $T$.  From the shape of the curves we can safely say 
that, for our sample, pinning due to surface barriers does not play 

\begin{table}
\caption{Experimental fitting parameters obtained from the 
         relaxation data in the low temperature regime ($T \protect\lesssim 
         20\,{\mathrm{K}}$) and in the high temperature regime ($T 
         \protect\gtrsim 40\,{\mathrm{K}}$).} \protect\label{FitPara}
\begin{tabular}{lcccc}
     & $U_c$ (K) & $t_o$ (s) & $j_c(T=0) \, (\mathrm{A/cm^2})$ & $\mu$ \\ 
     \hline
     $T \lesssim 20$ (K) & 140 & $3 \cdot 10^{-2}$ & $1 \cdot 10^6$ & 
     $0$ \\
     $T \gtrsim 40$ (K) & 1000 & $3 \cdot 10^{-3}$ & ---  & 0.6 \\
\end{tabular}
\end{table}

\noindent a dominant role.  In thin superconducting strips of rectangular cross 
section, Meissner currents flow throughout the whole 
sample\cite{larkin72,huebner72} and not only in a surface layer of 
width $\lambda$.  Lorentz forces arising from the Meissner currents 
will therefore act on the vortices and give rise to a barrier of 
purely geometric origin.\cite{zeldov94}  This kind of geometrical 
barrier will {\it not\/} influence measurements performed in the 
``field off'' state, since Meissner currents do not flow in this state 
(the influence of the residual fields $H^{\mathrm{res}}$ is of minor 
importance, see Table \ref{ResField}).

\subsection{Current-Voltage Characteristics}

Inductive measurements of the relaxation of the remanent magnetization 
$M_{\mathrm{rem}}$ are a powerful tool\cite{zhukov,ries,paulVI} for 
the evaluation of $E(j)$ characteristics of a superconductor down to 
{\it very} low values of the electric field $E$.  The functional 
dependence between the electric field $E$ and the current density $j$ 
can be found as follows.  In the Coulomb gauge ${\bbox{\nabla \cdot 
A}} = 0$, the vector-potential ${\bbox{A}}$ can be expressed 
as

\begin{equation}
  {\bbox{A}}({\bbox{r}}) = \frac{1}{c} \, 
      \int \frac{{\bbox{j}}({\bbox{r'}})}{|{\bbox{r}}-{\bbox{r'}}|}
      \, d^3r' \, ,
\end{equation}

\noindent where ${\bbox{j}}$ is the current density.  For a disk-like 
geometry and for a constant current density ${\bbox{j}}({\bbox{r'}}) = 
j \, {\bbox{e_\phi}},~(j = \mathrm{const.})$, one obtains

\begin{equation} \label{vecpotA}
  A({\bbox{r}}) \simeq j \cdot \frac{1}{c} \, 
          \int_V \frac{d^3r'}{|{\bbox{r}}-{\bbox{r'}}|} \, ,
\end{equation}

\noindent which, together with the Faraday induction law 
$\partial_t {\bbox{A}} = -c \, {\bbox{E}}$, leads to

\begin{figure}
\includegraphics[width=0.95\linewidth]{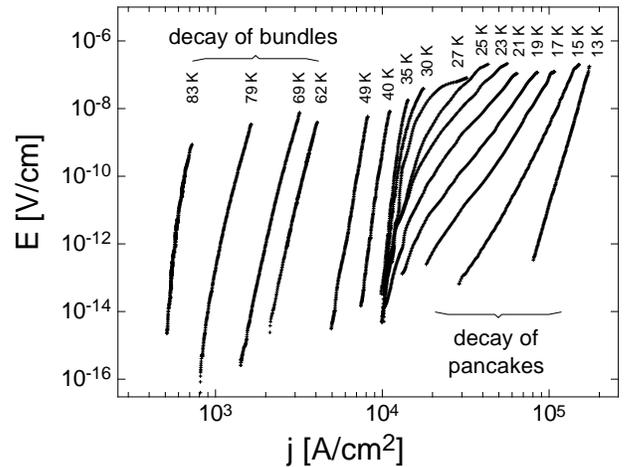}
\caption{Current-voltage characteristics as extracted from the 
         relaxation data for different temperatures $T$ at zero external 
         field $H$.}
\label{PicVI}
\end{figure}

\begin{equation} \label{Ej}
  E = \frac{\gamma}{c^2} h R \, 
      \frac{\partial j}{\partial t} ,~~~~\gamma \simeq 2 \, , 
\end{equation}

\noindent where $E$ is the mean electrical field in the sample, $R$ the 
sample radius and $h$ its thickness. The constant factor $\gamma$ 
arises from the assumption of a disk-like geometry for the sample.

The results obtained from the magnetic relaxation data and equation 
(\ref{Ej}) are plotted in Fig.\,\ref{PicVI}.  From the graph one can 
clearly distinguish the three regimes of vortex dynamics, which were 
previously discussed in this Section.  For current densities $j \le j_c$, 
the electric field $E$ due to the dissipative process of thermally 
activated vortex drift can be written as\cite{brandt94}

\begin{equation} \label{EU}
   E(j) = E_c \, \exp\left[ - U(j)/k_B T \right] \, .
\end{equation}

From Fig.\,\ref{PicVI} it is seen that for temperatures $T \lesssim 19 \, 
\mathrm{K}$ the $E(j)$ curves follow a power-law behavior.  As a matter 
of fact, inserting the logarithmic potential (\ref{U_log}) into formula 
(\ref{EU}) one obtains

\begin{equation} \label{EjPowerLaw}
   \frac{E}{E_c} = \left( \frac{j}{j_c} \right)^{U_c/k_B T} \, .
\end{equation}

From a fit to the curves in this temperature regime we obtain again $U_c 
\simeq 140 \, \mathrm{K}$, in agreement with the previous results.

\begin{table}
\caption{Values of $j_o$, $j_o / j_c(0)$, $L_c^c(0)$, $E_{\rm pc}$, 
         and $U_{\rm pc}$ as obtained from following formulas: $j_o = c \Phi_o 
         / (12 \protect\sqrt3 \pi^2 \lambda^2 \xi)$, $L_c^c \simeq \varepsilon 
         \xi (j_o / j_c)^{1/2}$, $E_{\rm pc} \approx \varepsilon_o d \, ( \xi / 
         \lambda )^2$, and $U_{\rm pc} \simeq \varepsilon_o d (j_c / j_o)$.  The 
         parameters used for the theoretical estimates are given for the 
         configuration where $H$ is perpendicular to the superconducting 
         layers.\protect\label{TheoPara}}
\begin{tabular}{lll}
     $j_o \approx 10^8 \, (\mathrm{A/cm^2})$ & $U_{\rm pc} \simeq 20$ (K) &
     $L_c^c(0) \simeq 2$ (\AA)  \\
     $j_o / j_c(0) \approx 100$ & $E_{\rm pc} \approx 3 \cdot 10^{-2}$ (K) \\ 
     \hline
     $\lambda_L \simeq 1800 \, (\mathrm{\AA})$ & $\lambda(0) = \lambda_L/\sqrt{2} 
     \simeq 1300 \, (\mathrm{\AA})$ & $d \simeq 
     15\,{(\mathrm{\AA})}$\\ 
     $\xi_{\rm BCS} \simeq 30 \, (\mathrm{\AA})$ & 
     $\xi(0) = \sqrt{0.54} \, \xi_{\rm BCS} \simeq 20 \, (\mathrm{\AA})$ &
     $\varepsilon \simeq 1/150$ \\
\end{tabular}
\end{table}

For temperatures $20 \, {\mathrm{K}} \lesssim T \lesssim 35 \, 
\mathrm{K}$, the electric field $E(j)$ behaves like a power-law only 
at high current densities.  For smaller values of the current density 
$j$, the different $E(j)$ curves tend to converge into the $T = 35 \, 
\mathrm{K}$ curve.  Further details about this temperature regime will 
be given in the next Section.

Finally, for temperatures $T \gtrsim 60 \, \mathrm{K}$, the $E(j)$ 
curves have a {\it negative\/} curvature in the $\log E$-$\log j$ 
plot, in agreement with the interpolation formula (\ref{U_interpol}) 
for the barrier as obtained from the WCPT.  The fitting 
parameter $U_c \approx 1000 \, \mathrm{K}$ is again consistent with 
our previous results.

\section{Summary and Discussion}\label{discussion}

The results obtained in the previous Section for the strongly layered 
Bi2212 single crystal in magnetic fields $H \perp ab$-planes are now
summarized and further discussed within the frame of WCPT. 

For temperatures $T \lesssim 20 \, {\mathrm{K}}$, the activation 
barrier for vortex motion $U(j)$ depends logarithmically on the 
current density, while the time relaxation of the current density $j$ 
follows a power-law behavior as given by formula (\ref{PowerLaw}).  
According to the discussion in Section \ref{TheoAspec}, an 
approximately logarithmic current dependence in the activation barrier 
(\ref{U_log}) is obtained within the single vortex pinning situation.  
However, in Section \ref{ExpAn} it has been shown that for 
temperatures $T \lesssim 20 \, {\mathrm{K}}$ and small enough magnetic 
fields, the correlation length along the $c$-axis $L_c^c$ is much 
smaller than the interlayer distance $d$.  This indicates that for 
this regime pinning involves elementary pancake-vortices.

On the other hand, within the most simple approach (see Section 
\ref{TheoAspec}), for decreasing current densities $j$ the activation 
barrier for single pancakes $U(j)$ is expected to be a constant, 
whereas the measured activation barrier is found to be logarithmic up 
to temperatures $T \simeq 19 \, {\mathrm{K}}$.  The non-constant 
behavior of the measured activation barrier $U(j)$ suggests that there 
are residual interactions which were not considered in the most simple 
approach and which lead to an increase of the elastic energy for 
decreasing $j$.  This argument is also supported by following 
considerations: The collective pinning energy for single 
pancakes,\cite{blatter} which is the relevant parameter for the 
determination of quantities such as the critical current density $j_c$ 
and the depinning energy, is given by $U_{\rm pc} \simeq \varepsilon_o 
d (j_c / j_o)$.  For the parameters of Table \ref{FitPara} and 
\ref{TheoPara} one finds that $U_{\rm pc} \simeq 20 \, \mathrm{K}$.  
Furthermore, the energy which is relevant for creep of 
pancake-vortices is expected to be bigger than $U_{\rm pc}$, but of 
the same order of magnitude.  However, this estimated energy is still 
small as compared to the values obtained for the activation energy 
$U(j)$ plotted in Fig.  \ref{PicULow}, indicating that for creep of 
pancake-vortices additional interactions have to be considered.  A 
possible idea leading to coupling of the pancake-vortices into an 
elastic plane\cite{blatter} for decreasing current densities $j$ is 
the concept of variable-range hopping.  As discussed in Section 
\ref{TheoPara}, due to the randomness in the energies of the 
metastable state, pancakes will hop over larger distances as the 
current density $j$ decreases.  The shear interaction energy $c_{66} d 
u^{2}$ will therefore grow with increasing hopping distance $u$ and 
for low enough current densities $c_{66} d u^{2}$ will become of the 
order of the pinning energy $U_{\rm pc}$.  In summary, the vortex 
system is expected to go over from a VRH regime (creep of individual 
pancake-vortices) to a 2D collective creep regime at low current 
densities.

\begin{figure}
\includegraphics[width=0.95\linewidth]{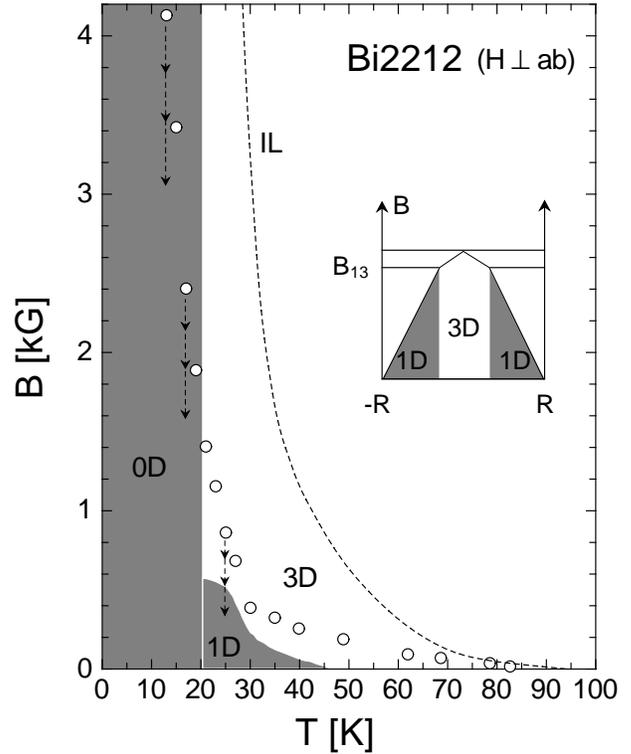}
\caption{Qualitative low field phase diagram of the vortex state in 
Bi2212 as proposed in Fig.\,\ref{PicPD_I} (The 2D regime is not 
shown in this plot since it is not relevant for the present data 
range).  The open circles in the graph are the estimated values of the 
$B$-field at the center of the sample ($B_{\rm center}$) at the 
starting time $t = t_s \simeq 18 \cdot 10^{-3} \, \mathrm{s}$, just 
after removing the external field $H$.  The vertical arrows 
qualitatively show the time evolution of $B_{\rm center}$ during the 
relaxation of the remanent magnetization $M_{\mathrm{rem}}$.  The 
insert represents schematically the profiles of the induction $B$ in 
the remanent state for temperatures between $20 \, {\mathrm{K}}$ and 
$40 \,{\mathrm{K}}$.  The shaded areas indicate the different pinning 
regimes throughout the sample (here $R$ is the sample radius).  In 
particular, the parameter $B_{13}$ in the insert indicates the value of 
the induction $B$ in the sample at which 1D pinning goes over into 3D 
pinning.  The smoothed irreversibility line IL in the graph was 
obtained from the data of Schilling {\it et al.\/} 
\protect\cite{schilling} } \label{PicPD}
\end{figure}

At higher temperatures ($T \gtrsim 20 \, {\mathrm{K}}$), where bulk 
pinning becomes relatively weak, barriers arising from the geometry of 
the sample and from surface effects can play an important role.  As 
shown in Section \ref{ExpAn}, in the low magnetic induction regime 
(``field off'' creep measurements) and for our specimen, these 
contributions are of minor importance as compared to the contributions 
of bulk pinning.

For temperatures $T \gtrsim 40 \, {\mathrm{K}}$, the activation 
barrier $U(j)$ is found to follow a power-law behavior 
(Fig.\,\ref{PicMaley_High_Temp}) that can be accurately fitted with 
formula (\ref{U_interpol}).  Within WCPT it has been shown that a 
potential of the form of (\ref{U_interpol}) leads to a time relaxation 
of the current density $j$ as given by the interpolation formula 
(\ref{j_interpol}).  From the fit to the $T = 69 \, {\mathrm{K}}$ data 
in Fig.\,\ref{PicDecayFit}, it is seen that for this temperature 
regime the time dependence of the current density $j$ is actually well 
described by formula (\ref{j_interpol}).  According to WCPT, a 
power-law potential with an exponent $\mu \simeq 0.6$, as obtained 
from the fits, indicates a regime of large bundle pinning.  A list of 
the obtained fitting parameters is given in Table \ref{FitPara}.

No unique functional dependence between $U$ and $j$ could be found for 
the temperature range $20 \, {\mathrm{K}} \lesssim T \lesssim 40 \, 
{\mathrm{K}}$.  Motivated by the present experimental results, the 
very low field regime of the Bi2212 pinning diagram has been 
investigated.  Fig.\,\ref{PicPD_I} shows the qualitative diagram 
obtained for this regime.  For temperatures $T$ above $T_{0}$, two 
different pinning regimes separated by $B_{13}$ are found involving 
vortex-pancakes and vortex-segments.  In order to describe the 
relaxation data in this temperature regime it is therefore necessary 
to estimate the induction in the center of the sample ($B_{\rm 
center}$) at the time $t = t_s \simeq 18 \cdot 10^{-3} \, \mathrm{s}$, 
immediately after removing the external field $H$.  These values are 
given for all temperatures by the empty circles in Fig.\,\ref{PicPD}, 
where the arrows indicate the time evolution of $B_{\rm center}$ 
during the relaxation of the remanent magnetization 
$M_{\mathrm{rem}}$.  For temperatures $T \lesssim 20 \, {\mathrm{K}}$ 
and $T \gtrsim 40 \, {\mathrm{K}}$, it follows that the whole sample 
is characterized by 0D and by 3D pinning, respectively.  On the other 
hand, as shown in the insert of Fig.\,\ref{PicPD}, for temperatures 
$20 \, {\mathrm{K}} \lesssim T \lesssim 40 \, {\mathrm{K}}$, the 
values of the initial field profile in the sample lead to the 
\textit{coexistence} of two different pinning regimes: 3D pinning in 
the central region of the sample and 1D pinning close to the borders.  
The simultaneous occurrence of two different pinning regimes may then 
provide an explanation why no simple functional dependence between $U$ 
and $j$ could be found for this temperature interval.

From the analysis of the relaxation data we find that vortex bundles 
are strongly pinned against thermal activation.  Nevertheless, due to 
their large size they can only sustain low flux density gradients, 
which means low critical current densities $j_c$.  On the other hand, 
vortex strings are weakly pinned, but being small in size they can 
sustain relatively high flux density gradients.  Much higher creep 
rates are therefore expected for collectively pinned vortex lines than 
for large vortex bundles.  Thus, a possible interpretation of the 
vortex dynamics observed for the temperature interval $20 \, 
{\mathrm{K}} \lesssim T \lesssim 40 \, {\mathrm{K}}$ is the following: 
The high relaxation rates which are measured at times $t \lesssim 1 \, 
{\mathrm{s}}$ (see Fig.\,\ref{PicMremNorm}) are mainly the result of 
the strong decay of the flux in the 1D regime at the border areas of 
the sample.  At times $t \gtrsim 1 \, {\mathrm{s}}$, most of the flux 
has left the sample and only a low flux density gradient of vortex 
strings remains.  As discussed in Section \ref{TheoAspec}, 
at low current densities $j$ the activated motion of vortex-lines 
involves hops of larger vortex segments by longer distances.  The weak 
relaxation rates of the current density $j$ measured at times $t 
\gtrsim 1 \, {\mathrm{s}}$ are then explained by the growth of the 
elastic energy for the activated motion of vortex strings at low 
current densities.

A feature that has recently attracted a lot of interest in the 
literature is the observation of a second peak in the magnetization 
loop.\cite{daeumling90} In our Bi2212 sample, as well as in several 
other works on Bi2212,\cite{zeldov95,kopylov90,yang,tamegai,yeshurun} 
this second peak is seen for magnetic inductions $B$ of the order of 
$\Phi_o / \lambda^2$ and for temperatures between $20 \, {\mathrm{K}}$ 
and $40 \, {\mathrm{K}}$.  As discussed in Section \ref{Vortex 
lattice}, for magnetic inductions $B \lesssim \Phi_o / \lambda^2$ the 
shear modulus $c_{66}$ starts to decrease exponentially so that, at low 
fields and for temperatures $T \gtrsim 20 \, {\mathrm{K}}$, a 1D 
pinning regime can arise (see Fig.\,\ref{PicPD_I}).  For the ascending 
branch of a magnetization loop it follows that, for fields larger than 
the field of first flux penetration $H_p$, the sample is expected to 
first enter into the 1D regime before gradually going over into 3D.  
As previously discussed, at a fixed temperature $T$, the value of 
$j_c$ is bigger for the 1D regime than for the 3D regime.  However, 
since the flux in the 1D regime has a much faster creep rate as 
compared to the 3D regime, it is very important to consider the 
\textit{time scale} for the measurement of the magnetization loop.  
For instance, if the time scale were very short ($t \to 0$), the 
effects of creep would be negligible and for an increasing 
(decreasing) magnetic field $H$ one would expect to measure a {\it 
decrease} ({\it increase}) in the magnetization as soon as the 
magnetic induction $B$ is of the order of $\Phi_o / \lambda^2$, where 
the flux in the sample goes over from 1D to 3D.  On the other hand, on 
the typical time scale of the measurement of a magnetization loop, the 
flux in the 1D regime is already strongly relaxed while the flux in 
the 3D regime is still close to its configuration in the critical 
state.  The value of the magnetization $M$ may then turn out to be 
smaller in the 1D regime than in the 3D regime, leading to the 
characteristic double peak in the magnetization loop as measured for 
temperatures between $20 \, {\mathrm{K}}$ and $40 \, {\mathrm{K}}$.

This interpretation is in agreement with previous reports (Refs.\ 
\onlinecite{yeshurun,chiku,elbaum}) which relate the second peak of 
the magnetization loop to the slower magnetization decay for the field 
range where the peak is observed.  Moreover, it is in agreement with 
the results of Ref.\ \onlinecite{zeldov95} regarding local induction 
measurements on a Bi2212 sample.  In the descending branch of the 
magnetization loop, at $T = 24 \, {\mathrm{K}}$ and for field values 
between $330 \, {\mathrm{Oe}}$ and $260 \, {\mathrm{Oe}}$, a change of 
slope of the field profile $d B_z(x)/d x$ is observed\cite{zeldov95} 
occurring at various locations inside the crystal starting from the 
edge regions and moving towards the center as the applied field is 
decreased (with $B_z$ being the induction parallel to the 
crystallographic $c$-axis).  Within the presented low field phase 
diagram of Bi2212, this change of slope is expected to occur at the 
crossover field $B_{13}$.

In conclusion, for temperatures $20 \, {\mathrm{K}} \lesssim T 
\lesssim 40 \, {\mathrm{K}}$, the observation of the second peak in the 
magnetization loop and of the high relaxation rates of the current 
density $j$ for times $t \lesssim 1 \, \mathrm{s}$ can both be related to 
the coexistence of two different pinning regimes inside the sample and 
to the strong difference in their relaxation rates.

\section{Acknowledgments}

It is a pleasure to acknowledge many helpful discussions with 
V.B.~Geshkenbein, C.~de~Morais-Smith, T.~Teruzzi, K.~Aupke, 
R.~Frassanito, and A.~Amann.  We are grateful to V.N.~Zavaritzky for 
providing the Bi2212 sample.  A.S.  would like to thank D.~Brinkmann 
for his kind support.  This work was supported by the Schweizerischer 
Nationalfonds zur F\"orderung der wissenschaftlichen Forschung and by 
the Eidgen\"ossische Stiftung zur F\"orderung der schweizerischen 
Volkswirtschaft durch wissenschaftliche Forschung.

\end{multicols}

\end{document}